\documentclass[a4paper,conference]{IEEEtran}

\usepackage[utf8]{inputenc}
\usepackage{graphicx}
\usepackage{hyperref}

\hypersetup{
  pdfauthor = {Guilherme Amadio and Benda Xu},
  colorlinks, urlcolor=black, citecolor=black,
  pdftitle = {Portage: Bringing Hackers' Wisdom to Science},
  pdfsubject = {Third International Workshop on HPC User Support Tools}
}

\title{Portage: Bringing Hackers' Wisdom to Science}

\author{
  \IEEEauthorblockN{Guilherme Amadio}
  \IEEEauthorblockA{São Paulo State University, Brazil\\
                    Gentoo Linux\\amadio@gentoo.org}
  \and
  \IEEEauthorblockN{Benda Xu}
  \IEEEauthorblockA{The University of Tokyo, Japan\\
                    Gentoo Linux\\heroxbd@gentoo.org}
}

\begin{document}
\maketitle

\begin{abstract}

Providing users of HPC systems with a wide variety of up to date software
packages is a challenging task. Large software stacks built from source
are difficult to manage, requiring powerful package management tools.
The Portage package manager from Gentoo is a highly flexible tool that
offers a mature solution to this otherwise daunting task. The Gentoo
Prefix project develops and maintains a way of installing Gentoo systems
in non-standard locations, bringing the virtues of Gentoo to other
operating systems. Here we demonstrate how a Gentoo Prefix
installation can be used to cross compile software packages for the
Intel® Xeon Phi™ known as Knights Corner, as well as to manage large
software stacks in HPC environments.

\end{abstract}

\section{Introduction}

Building and maintaining a large software stack on an HPC environment
requires powerful packaging tools. Packages provided by the core operating
systems from HPC vendors often lag significantly behind the latest
upstream releases, while users want to make use of new features in
recent software releases, or build against specific versions of software
packages for compatibility or consistency reasons. In order to satisfy
users, the package manager must take into account the multitude of
configuration options for each package as well as ensure that the build
order is correct. It must also be able to distinguish between build
time, run time, and post build dependencies, and be able to force
package rebuilds when configuration options of dependencies change,
without rebuilding the full tree of packages in the process. Finally,
due to the particularities of each HPC environment, it is highly
desirable for the package manager to offer the ability to easily apply
patches to specific software packages. Fortunately,
Gentoo's~\cite{gentoo} Portage~\cite{gentoo:portage} package manager is
able to do all this and much more. On HPC systems where users usually do
not have permission to install software into standard locations, Gentoo
Prefix \cite{gentoo:prefix} allows installations of a Gentoo environment
into non-standard paths.

\subsection{Gentoo Linux}

Gentoo Linux \cite{gentoo} is a source-based distribution with powerful
tools that make it an excellent development environment. Gentoo has been
first released by its founder Daniel Robbins in December of 1999, and
has been in active development since then. It is a metadistribution, in
the sense that it provides the necessary tools for the user to build
his[her] own customized version of Gentoo. Gentoo uses its own package
manager, Portage~\cite{gentoo:portage}, written in Python and inspired
by the ports system from FreeBSD. Gentoo is not based on any Linux
distribution, but is itself the root of popular distributions like
Sabayon Linux, and innovative operating systems such as
CoreOS \cite{coreos} and Google's ChromeOS \cite{chromiumos}.

\subsection{Gentoo Prefix Project}
\label{sec:portage}

The Gentoo Prefix project \cite{gentoo:prefix} brings the virtues of
Gentoo Linux---such as high configurability, performance tuning, and
automated package dependency management---to different operating
systems. This is useful for installing software into systems where the
user might not have administrator privileges, or simply to use Gentoo's
Portage package manager to automate the process of fetching source code,
building, and installing software packages available in Gentoo and
having their dependencies automatically handled, including the
configuration options of each package.

Gentoo Prefix uses the host system's kernel and standard C library, but
all other tools are installed and managed during the bootstrapping of
the system. On Linux, Gentoo Prefix can also compile its own standard C
library~\cite{gentoo:rap}. Installation of the Gentoo Prefix environment
is performed by a shell script that installs the Portage package manager
into a temporary location, then uses it to bootstrap a compiler and
install the base system, which allows the user to compile and install
other sofware packages available in the Portage database. Gentoo Prefix
can be installed not only on Linux, but also on Mac~OS~X, BSD, AIX,
Solaris, etc. On Mac~OS~X, Gentoo Prefix is similar to projects
such as Homebrew, MacPorts (source based) and Fink (binary based).

\subsection{Portage Package Manager}
\label{sec:ebuild}

Portage is a GPLv2 package management system based on FreeBSD's ports
collection. Portage consists of two main parts, the \emph{ebuild}
system, and the \emph{emerge} command line utility. The ebuild system is
responsible for executing build instructions and installing packages,
while emerge provides dependency management, and servers as the
interface to ebuild. The relationship between ebuild and emerge is
similar to that of rpm and yum on Red Hat, and that of dpkg and apt on
Debian.

Portage packages are special bash shell scripts called
\emph{ebuilds}~(not to be confused with the ebuild tool itself that is
used to run them). Ebuilds are similar to spec files in SRPMs; they
contain information on how to download, configure, compile, and
install software, as well as their dependency requirements. Gentoo's
main ebuild repository has more than 27000 ebuilds available for a
variety of architectures. Additional packages are available via
official and unofficial package overlays that complement the main tree
of packages.  The special syntax used in ebuild scripts is
standardized in the Package Manager Specification (PMS)
\cite{gentoo:pms} document. The PMS documents the behavior of Portage
so that Gentoo packages can be managed by alternative package
managers, such as paludis \cite{paludis} and pkgcore \cite{pkgcore}.

The Portage system has the concept of \emph{USE flags} to allow the
user to configure compile-time options of software packages. USE flags
affect which dependencies are required to build a package, which
allows, for example, a headless server to be installed with a lighter
system footprint, by stripping all options for building a graphical
environment. USE flags can be used in HPC systems to enable or disable
MPI support, to choose which version of the Python interpreter a
module should target, to enable specific instruction sets for a given
architecture, among many others.

For software packages that require specific versions, Portage has the
concept of package \emph{SLOTs} \cite[p.~27]{gentoo:pms}, which allow multiple versions of a same
package to be installed simultaneously on the same system. This is very
useful for, e.g., installing packages that depend on incompatible
versions of the same library, or to have multiple compiler versions
available for development and testing. SLOTs can also be used in package
dependencies to indicate when a package should be rebuilt should the
installed version of its dependencies change.


\section{Gentoo Prefix on an HPC environment}
As Gentoo Prefix runs on all the GNU/Linux Distributions, it supports
GNU/Linux-based HPC systems off-the-shelf. It is also reported that
Gentoo Prefix runs on Cray with some minor site-specific modifications
\cite{cray}.

Gentoo Prefix uses ebuilds from Gentoo's main repository~(Section
\ref{sec:ebuild}) and receives cutting edge updates by hundreds of
developers and thousands of user contributors. It is a general-purpose
and full-featured environment, which mitigates a big portion of the
complexity and workload of HPC software maintenance. \LaTeX{} and
pandoc for publication, emacs and vim for editing, vnc and xpra for
remote display, to name a few, are all available.

From the users' point of view, Gentoo Prefix vastly expands the tools
available for HPC. From the administrators' point of view, Gentoo
Prefix supports a diversity of user needs and at the same time keeps the
core operating system stable and secure.

Installation instructions for Gentoo Prefix can be found on the project
website \cite{gentoo:prefix}. The \texttt{bootstrap-prefix.sh} script
automates most of the process (\texttt{\$} shown below is the normal
user command line prompt).
\begin{verbatim}
(download script from project website)
$ chmod 755 bootstrap-prefix.sh
$ ./bootstrap-prefix.sh
(follow the instructions)
\end{verbatim}

The bootstrap process consists roughly of three stages: installing a
temporary version of Portage, boostrapping a compiler using the host's
toolchain, then bootstrapping a full system with the bootstrapped
compiler. On Linux, a C standard library can also be part of the
bootstrap process, in which case the kernel is the only piece shared
between the host and the prefix environment.  To have the C standard
library from Gentoo, rename the script before invoking it.
\begin{verbatim}
$ mv bootstrap-prefix.sh bootstrap-rap.sh
$ ./bootstrap-rap.sh
(follow the instructions)
\end{verbatim}

The script works as an interactive installer; it will ask the user for
some information and discuss options based on the system being
bootstrapped on. Once the script finishes running, it will create a
\texttt{startprefix} script at the root of the bootstrapped system.
Running that script brings the user into the prefix environment, where
\texttt{emerge} and other tools become available to let the user install
more packages.
\begin{verbatim}
$ ${EPREFIX}/startprefix
Entering Gentoo Prefix /micfs/gentoo
$
\end{verbatim}

The following sections present several case studies to show the
flexibility of Gentoo Prefix.

\subsection{Nested Job-Scheduling with Slurm}
\label{sec:slurm}
Although HPC programs are typically parallel, for some embarrassingly
parallel tasks like data analysis and simulation, it is enough to have
a grid infrastructure running many single-threaded workers divided by
input data.

Some supercomputing site allocates resources in units of nodes and
leaves finer scheduling within nodes to users. For MPI-based parallel
programs it is straight-forward, but for single-threaded programs it
is tricky to efficiently and reliably use all the CPUs in an allocated
node.

Instead of writing \textit{ad hoc} shell scripts to launch multiple
single-threaded programs, slurm or torque available in Gentoo could be
used as a nested scheduler and allocate jobs in units of
CPUs.

\begin{figure}[htb]
  \centering
  \includegraphics[width=0.5\textwidth]{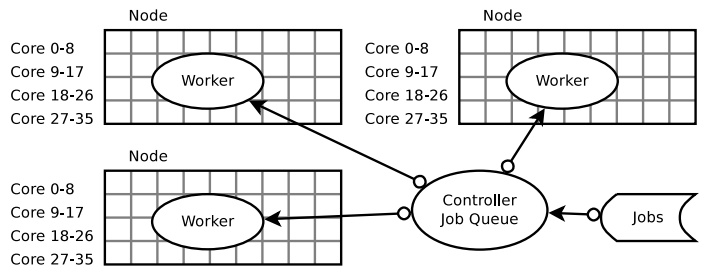}
  \caption{User slurm is nested in site PBS Pro. The nodes are
    allocated by PBS and each slurm worker runs on each node.
    Single-threaded jobs are controlled and queued by user slurm
    controller. Resources within a node, such as CPUs and memory, are
    individually allocated.}
  \label{fig:slurm}
\end{figure}

Assuming that the host scheduler is PBS Professional and has been
installed by the site administrators, a nested slurm
scheduler~(Fig.~\ref{fig:slurm}) can be installed by a normal user with
Gentoo Prefix on a filesystem shared between all nodes.
\begin{verbatim} 
$ emerge sys-cluster/slurm 
\end{verbatim}

After editing \texttt{EPREFIX/etc/slurm/slurm.conf} to enable individual
allocation of resources within a node, the slurm worker can be launched by
\texttt{slurmd} and the slurm controller by \texttt{slurmctld}.
\texttt{EPREFIX} is the installation offset of Gentoo Prefix.

It is recommended to use munge for authentication between
\texttt{slurmd} and \texttt{slurmctld}. OpenRC, a dependency aware
service manager, is suitable to manage the relation between munge and
slurm daemons. We create named runlevels for the slurm worker and controller,
\begin{verbatim}
$ mkdir EPREFIX/etc/runlevels/worker
$ mkdir EPREFIX/etc/runlevels/controller
\end{verbatim}
add services to the corresponding runlevels,
\begin{verbatim}
$ rc-update add munged controller worker
$ rc-update add slurmctld controller
$ rc-update add slurmd controller
\end{verbatim}
and start slurmctld on the controller node (or login node) with
\begin{verbatim}
$ openrc controller
 * Starting munged ...
 * Starting slurm control daemon ...
\end{verbatim}
Then, we submit a PBS batch script to launch slurm workers,
\begin{verbatim}
$ cat slurmd.sh
#!/bin/sh
#PBS -l select=32:ncpus=36:mpiprocs=1
#PBS -l walltime=48:00:00
source EPREFIX/etc/profile
mpirun sh -c "openrc worker; sleep 2d"
$ qsub slurmd.sh
\end{verbatim}
In the script above, 32 nodes, each having 36 CPUs, are requested for 2
days. \texttt{mpirun} is used to launch 1 slurm worker per node.

After the workers are started, single-threaded jobs can be submitted
with slurm tools.

\subsection{Memory Profile with Valgrind and massif-visualizer}
\label{sec:massif}
In the context of HPC, large scale problems could become
memory-bounded, for example when fill-in happens to a sparse matrix
based algorithm. Valgrind is traditionally a memory debugging tool
for memory leak detection. A heap profiler, massif~\cite{massif}, is
also distributed with valgrind.
massif-visualizer~\cite{massif:visualizer} is a handful GUI visualizer
for interpreting massif outputs besides the default CLI-based
\texttt{ms-print}.

massif-visualizer is a KDE application. It depends on a recent KDE
framework, which is a fairly large and complex software stack. Gentoo
Prefix readily provides the whole KDE stack thanks to the shared
\textit{ebuilds} with the main Gentoo repository. As massif-visualizer
is already in the KDE overlay, building it is straightforward,
\begin{verbatim}
$ layman --add=kde
$ emerge kde-misc/massif-visualizer
\end{verbatim}
and of course in addition to valgrind
\begin{verbatim}
$ emerge dev-util/valgrind
\end{verbatim}

To resolve the function names in the stack in valgrind output, debug
symbols are needed. A neat feature is provided by Portage to control
build environment in a per package-basis. For example, to get memory
profile of a quantile regression package \texttt{quantreg} of GNU R,
debug symbols of glibc, R and quantreg are needed. In
\texttt{EPREFIX/etc/portage/env/debug-cflags.conf}, add
\begin{verbatim}
CFLAGS="-O2 -ggdb -pipe"
CXXFLAGS="${CFLAGS}"
FEATURES="${FEATURES} splitdebug"
\end{verbatim}
and in \texttt{EPREFIX/etc/portage/package.env}, add
\begin{verbatim}
sci-CRAN/quantreg debug-cflags.conf
dev-lang/R debug-cflags.conf
sys-libs/glibc debug-cflags.conf
\end{verbatim}
and rebuild the packages,
\begin{verbatim}
$ emerge sys-libs/glibc dev-lang/R \
> sci-CRAN/quantreg
\end{verbatim}

To produce and view the memory profile,
\begin{verbatim}
R -d "valgrind --tool=massif \
> --massif-out-file=rq.prof" -f rq.R
$ massif-visualizer rq.prof
\end{verbatim}
where \texttt{rq.R} is the R script to be profiled, its content being
irrelavent to this discussion.

\subsection{Cross Compiling Software for the Xeon Phi™}

Portage can also be configured to cross compile software for different
platforms. The usual way of doing so is to follow the Gentoo Embedded
Handbook \cite{gentoo:embedded} to set up the appropriate GCC toolchain
and setup the right Portage profile. However, cross compiling for the
Xeon Phi™ is somewhat different, as only the Intel® C/C++ Compiler has
full support for the Xeon Phi™'s IMCI (Initial ManyCore Instructions)
architecture, which, despite being quite similar to the x86
architecture, has some subtle differences. The fact that Knights Corner
Xeon Phi™ cards are used as coprocessors and their boot process does not
follow a conventional procedure is irrelevant for cross compiling. Once
the Xeon Phi™ is properly set up, it basically works as a headless
server; users can just ssh into it and run their programs. In our setup,
a server with CentOS 7 and Intel tools installed was used to host the
Gentoo Prefix installation. The host system also needs to have both the
Intel® C/C++ compiler and Intel®'s Manycore Platform Software Stack
(MPSS) in order to fulfill the requirements for cross compiling.

Since Python is a popular language for scientific computing in general,
we demonstrate how to cross compile Python 2.7 for the Xeon Phi™. The
first step was to create a root directory for Portage, where all the
configuration files reside, and where all software packages are
installed.

In our system, we used a separate partition (\verb|/micfs|) to hold the
Gentoo Prefix (\verb|/micfs/gentoo|) and the root of the tree of cross
compiled software (\verb|/micfs/gentoo-mic|). The directory
\verb|/micfs/gentoo-mic/etc/portage| contains all files needed to
configure Portage to use the Intel compiler, and the proper compilation
flags to cross compile for the Xeon Phi.
The first file needed is a \verb|bashrc|:
\begin{verbatim}
$ cd /micfs/gentoo-mic/etc/portage
$ cat > bashrc <<-EOF
. /opt/intel/bin/compilervars.sh intel64
EOF
\end{verbatim}
This file is sourced by Portage, which gains access to \verb|icc|
by placing it in its \verb|$PATH|.

Now, the most important file, \verb|make.conf|, is used to properly
configure the compiler to produce binaries for the Xeon Phi™:

{\footnotesize
\begin{verbatim}
$ cat > make.conf <<-EOF
ACCEPT_KEYWORDS="~amd64-linux"

CONFIG_SHELL="/micfs/gentoo/bin/bash"

CBUILD="x86_64-pc-linux-gnu"
CHOST="x86_64-k1om-linux-gnu"

CC=icc
CXX=icpc
MPI_C=mpiicc
MPI_CXX=mpiicpc
LD=xild
AR=xiar

CFLAGS="${CFLAGS} -mmic -O3"
CXXFLAGS="${CXXFLAGS} -mmic -O3"
FFLAGS="${FFLAGS} -mmic -O3"

# append link flags, order is important!
# we inject RPATHs to avoid setting
# LD_LIBRARY_PATH later
LDFLAGS="${LDFLAGS} -mmic \
         -L/opt/intel/lib/mic \
         -Wl,--rpath,/opt/intel/lib/mic \
         -L/micfs/gentoo-mic/usr/lib \
         -Wl,--rpath,/micfs/gentoo-mic/usr/lib \
         -limf -lsvml -lirng -lintlc"

USE_PYTHON="2.7"
PYTHON_TARGETS="python2_7"

USE="-* unicode python ncurses readline threads xml"
EOF
\end{verbatim}
}

To cross compile Python 2.7 \cite{python-k1om-youtube}, a patch
\cite{libffi-k1om} needs to be applied to one of its dependencies,
libffi, in order for it to properly cross compile for the Xeon Phi™.
Portage supports user patches to be applied by simply dropping them into
\verb|EPREFIX/etc/portage/patches| in a subdirectory corresponding to
the package to be patched, with an optional version. In this case, we
dropped the patch file \verb|libffi-3.2.1-k1om.patch| into the
subdirectory \verb|dev-libs/libffi-3.2.1| under
\verb|/micfs/gentoo-mic/etc/portage/patches|. If the version is
ommitted, the patch will apply to all versions.

After everything has been properly configured as shown above, Python
and all its dependencies can be cross compiled with a single command
\begin{verbatim}
$ emerge \
> --root=/micfs/gentoo-mic \
> --config-root=/micfs/gentoo-mic \
> --prefix=/ python:2.7
\end{verbatim}

The command above installs in total 26 packages, including Python.
To confirm that the software was properly cross compiled, we can check
the architecture of the generated libraries with \verb|readelf|:
\begin{verbatim}
$ readelf -h \
> /micfs/gentoo-mic/usr/bin/Python2.7 |
> grep Machine
  Machine:              Intel K1OM
\end{verbatim}

Python modules can now be installed with emerge too. Due to
incompatibilities, not all software can be cross compiled this way.
Python 3.5, for example, needs further patching before it can be
installed by Portage. However, for basic software like \verb|bash|,
\verb|rsync|, and many other tools that facilitate the users' life
when logged in on the Xeon Phi™, Portage works quite well.

\section{Assessment of Portage for HPC}

With more than 10 years development, Portage has become an excellent
and mature package management system. It is general purpose and used
in a diversity of scenarios. Nevertheless, it is still worthwhile to
assess whether Portage meets the specific needs of the HPC
community. The technical report from the software packaging working
group of HEP software fundation~\cite{hsf:package} has summarized the
requirements of HPC from the point of view of high energy physics,
which is also applicable to other fields.

As the core of Gentoo is a meta distribution, Portage naturally satisfies
all the requirements:

\begin{enumerate}
\item Portability: Portage runs on GNU/Linux, Android/Linux, Solaris,
  AIX, Mac OS X, Windows and more (Section \ref{sec:portage}).
\item Multiple versions: Portage supports building and installing
  multiple stacks in parallel with the SLOT mechanism
  (Section \ref{sec:ebuild}).
\item Ease: Portage is self-contained, ebuilds for new packages are
  easy to add (Section \ref{sec:portage}), and the bootstraping process
  on POSIX environments is simple.
\item Support: The Gentoo community is friendly, active and
  supportive. Documentaion of Gentoo is abundant and informative.
\item Performance: Gentoo is a Linux distribution famous for keeping
  users' CPU warm. Compiling thousands of packages is a daily exercise
  of an average Gentoo user. Portage provides a mature build system
  scrutinized by hackers all around the world and tested in every
  extreme setups. The vcs support (checking out sources from cvs, svn,
  git, mercury, etc.) implemented by eclasses is excellent.
\end{enumerate}

\section{A Deeper Look at Multiple Versions and Build Configurations}

Both the HSF technical report~\cite{hsf:package} and Gamblin et
al.~\cite{spack} argue that multiple versions and multiple build
configurations of the same package should be available at the same
time to support wide user groups in a shared HPC environment. Gamblin
et al.~\cite{spack} also put forward explicitly that software build is
by its nature combinatorial, therefore a package manager should
support all the combinations natively. Although from an administrator's
point of view the system can potentially have all kinds of combinations,
for a single user or a closely collaborated group this is an overkill.

If each user group maintains their own software stack, support for
multiple versions and build configurations is rarely necessary.
Therefore the advantage provided by multiple versions and build
configurations is limited to administrators.

Only a few packages need explicit version tracking and specific build
configurations, like compilers/interpreters (gcc-4 vs. 5 vs. 6,
python-2 vs. 3, ruby-2.2 vs. 2.3), MPI implementations, blas/lapack
implementations, autoconf/automake versions, complete rewrites of
frameworks (Geant-3 vs. 4). For the majority of relatively small tools
the cost of complexity exceeds the benefit of diversity of multiple
versions and build configurations. Ultimately it is a balance between
simplicity and diversity. Portage provides straightforward solutions
for installing multiple package versions with \emph{SLOTs} and for
selecting among alternative implementations with \emph{eselect}, or
disabling all implementations by default and only enabling them after
conscious and educated decisions.  Conversely, installing a multitude
of versions and build configurations of a single package \emph{by
  default} is pursuing diversity to the extreme.

Project management in scientific software is far from mature. Authors
of scientific packages often value instant gratification (publish of
an article) more than long term maintainability (robust build system,
stable API, etc.). ABI and even API incompatibilities between minor
updates are common. Function calls from libraries could give different
results when linked against different versions of stardard c++
runtime.  It therefore becomes crucial for a package manager to
support multiple versions and build configurations to facilitate
reproducibility in research. Package managers with such a feature
provide an instant solution to the chaos of scientific software like a
painkiller.  Providing such a feature by default may encourage bad
practice.

There is a growing consensus that computer programs are replacing
mathematics to become the most used language of science. There will be
more sophisticated scientific software in the future. The value of
good practice in project management will be more conspicuous, most of
which are already established in free and open source software
communities.  In the context of HPC, it is more important to draw on
the experience of software communities than to circumvent issues
caused by bad designs.

Therefore supporting multiple versions and build configurations by
default calls for more caution, especially for a healthy scientific
ecosystem in the a long run.

\section{Conclusion}
\label{sec:conclusion}
Portage is a powerful package management system from Gentoo. It is
robust, mature, well-documented and user-friendly. It is especially
suitable in HPC setups. Gentoo Prefix is easy to bootstrap, and
easier to bootstrap a new instance from an existing one. It brings
the power of a well-received metadistribution to the users, and at the
same time reduces the heavy maintanance burden of administrators.

Portage and Gentoo repositories are scrutinized by hundreds of
developers worldwide, many of whom are software professionals and
respected hackers. They grow out of a group of people who engineer
complex software. Their wisdom to handle complexity is invaluable to
scientific communities.

Portage is a perfect bridge between hackers and scientists, and gives
an ultimate solution to the chaotic \emph{status quo} of HPC software
stacks.



\end{document}